\documentclass[12pt]{article}
\usepackage{latexsym}

\def\bs{\begin{subequations}}
\def\es{\end{subequations}}

\catcode`\@=11

\newtoks\@stequation
\def\subequations{\refstepcounter{equation}
  \edef\@savedequation{\the\c@equation}%
  \@stequation=\expandafter{\theequation}%   %only want \theequation
  \edef\@savedtheequation{\the\@stequation}% % expanded once
  \edef\oldtheequation{\theequation}%
  \setcounter{equation}{0}%
  \def\theequation{\oldtheequation\alph{equation}}}

\def\endsubequations{\setcounter{equation}{\@savedequation}%
  \@stequation=\expandafter{\@savedtheequation}%
  \edef\theequation{\the\@stequation}\global\@ignoretrue}

\makeatletter%  allow access to internal LaTeX commands
        \renewcommand{\theequation}{\thesection.\arabic{equation}}%
        \@addtoreset{equation}{section}%
\makeatother%  turn off access to internal LaTeX commands

\renewcommand{\thefootnote}{\fnsymbol{footnote}}

\begin{document}

\begin{titlepage}

March 20, 2017

\begin{center}        \hfill   \\
            \hfill     \\
                                \hfill   \\

\vskip .25in

{\large \bf Numerical Method in Classical Dynamics \\}

\vskip 0.3in

Charles Schwartz\footnote{E-mail: schwartz@physics.berkeley.edu}

\vskip 0.15in

{\em Department of Physics,
     University of California\\
     Berkeley, California 94720}
        
\end{center}

\vskip .3in

\vfill

\begin{abstract}

A set of algorithms is presented for efficient numerical 
calculation of the time evolution of classical dynamical systems.
Starting with a first approximation for solving the differential 
equations that has a ``reversible'' character, we show how to bootstrap 
easily to higher order accuracy. The method, first shown for a single particle in one dimension, is then neatly extended to many dimensions and many particles
\end{abstract}

\vfill


\end{titlepage}

\renewcommand{\thefootnote}{\arabic{footnote}}
\setcounter{footnote}{0}
\renewcommand{\thepage}{\arabic{page}}
\setcounter{page}{1}

\section{The Problem}%1

We start by considering Newton's Law of motion for one particle moving in one dimension; in the Appendix we show how to extend this method neatly to many dimensions and many particles.  
We write a pair of first order time evolution equations:
\begin{equation}
\frac{d}{dt} \left(\begin{array}{c} x(t)\\v(t) \end{array}\right) = 
\left(\begin{array}{c} v(t)\\f(x(t)) \end{array}\right) = 
M \left(\begin{array}{c} x(t)\\v(t) \end{array}\right)\label{a1}
\end{equation}
where both $x$ and $v$ are time dependent functions to be determined 
at some later time $t$, given their values at some initial time 
$t=0.$ The force is given by some specified function $f(x)$; and the 
quantity $M$ is defined as the (non-linear) matrix/operator 
specified above.

For simplicity I will write
\begin{equation}
\psi(t) = \left(\begin{array}{c} x(t)\\v(t) \end{array}\right).
\end{equation}
We assume that there exists an operator $E(tM)$ that is the 
exact propagator:
\begin{equation}
\psi(t) = E(tM)\;\psi(0),\;\;\;\;\;\;\;\;
E(tM) = \lim_{N \rightarrow \infty} (1+\frac{t}{N}M)^{N}.
\end{equation}
The addition property, $E(t_{1}M)E(t_{2}M) = 
E((t_{1}+t_{2})M)$ follows. Alternatively, we may write,
\begin{equation}
\frac{d}{dt} E(tM) = M\;E(tM).
\end{equation}

This formalism is familiar in the case where M is a general linear 
operator, and E is simply the ordinary exponential function; however,
it is also appropriate for non-linear operators, 
as derived in reference \cite{CS}.

Our objective is to show simple and accurate approximations to the 
operator $E(\delta M)$, for small time-steps $\delta$, for use in 
automated computations.

\section{The General Method}%2

Following the general method given in \cite{CS}, we start by 
constructing an approximate propagation operator $R(\delta)$ with the following 
properties:
\begin{equation}
R(\delta)\;R(-\delta) = 1,\label{b1}
\end{equation}
and
\begin{equation}
R(\delta)= E(\delta M + \delta^{3}X_{3} + \delta^{5}X_{5}+ \ldots).\label{b2}
\end{equation}

Rather than expanding the approximate result $\psi(t+\delta) \approx R(\delta)\;\psi(t)$ in a power series in 
$\delta$, we represent $R$ as the exact propagator for some other 
problem, which is expanded about the true one: based upon $M$.  The restriction 
(\ref{b1}) means that only odd powers of $\delta$ occur in the 
expansion (\ref{b2}).  The quantities $X_{3}$, $X_{5}$, etc., are 
unknown. Our method will show how to eliminate those higher order 
errors, step by step.

One more general property of the abstract propagator function $E$ is 
the following.
\begin{eqnarray}
E(A)E(B) = E(A+B + \frac{1}{2}[A/B] + \frac{1}{12}[(A-B)/[A/B]] 
\ldots), \\ \label{b3}
E(A)E(B)E(A) = E(2A+B - \frac{1}{6}[(A+B)/[A/B]] 
\ldots).
\end{eqnarray}
This is the nonlinear extension of the Baker-Campbell-Hausdorff theorem for the 
product of exponentials of non-commuting linear operators. The only 
difference is that, instead of the commutator $[A,B]=AB-BA$ for 
linear operators, we have the  ``slash commutator'' $[A/B]= A/B - 
B/A$ for nonlinear operators, as defined in reference \cite{CS}.

Now we proceed. The initial operator $R(\delta)$ is correct to order 
$\delta^{2}$ and so we call it $R_{2}(\delta)$. Now we construct the 
following sandwich:
\begin{equation}
R_{4}(\delta) = R_{2}(\beta \delta)\;R_{2}(\gamma 
\delta)\;R_{2}(\beta \delta);\label{b4}
\end{equation}
and try to choose the constants $\beta, \gamma$ so that
\begin{equation}
R_{4}(\delta) = E(\delta M + \delta^{5}Y_{5} + \ldots).\label{b5}
\end{equation}
By working with Equation (\ref{b3}) we find the simple requirements,
\begin{equation}
2\beta+\gamma = 1, \;\;\;\;\; 2\beta^{3}+\gamma^{3} = 0, \;\;\;\;\;
\beta = (2-2^{1/3})^{-1}, \;\;\;\;\; \gamma = 
-2^{1/3}\;\beta.\label{b6}
\end{equation}

This new formula (\ref{b4}) may be read as follows: Take a step 
forward of length $1.351207\ldots \delta$, then take a step backwards of length 
$1.702414\ldots \delta$, 
then another step forward of length $1.351207\ldots \delta$. The result will be one step 
forward of length $\delta$ - with errors of order $\delta^{5}$.

\section{Choosing R}%3

The real challenge now is to construct $R(\delta)$, seemingly accurate  only 
to first order in $\delta$ but restricted by the requirement 
(\ref{b1}).

Here is one suggestion, for the particular problem we started with 
(\ref{a1}), that is built in the ``sandwich'' manner.
\begin{eqnarray}
R_{2}(\delta) = D_{x}(\delta/2)D_{v}(\delta)D_{x}(\delta/2) \label{c1} \\
D_{x}(\delta) \left(\begin{array}{c} x \\ v \end{array}\right) = 
\left(\begin{array}{c} x+\delta v \\ v \end{array}\right) \label{c2} \\
D_{v}(\delta) \left(\begin{array}{c} x \\ v \end{array}\right) = 
\left(\begin{array}{c} x \\ v +\delta f(x) 
\end{array}\right).\label{c3}
\end{eqnarray}

It should be apparent that this formulation is very easy to program 
for automated computation. On the other hand, it is rather cumbersome if one 
writes out explicit formulas for the overall result of this sequence 
of operations.

\section{Numerical Examples}%4
I have applied this method to a simple problem, the Kepler orbit in a 
plane. With the initial conditions $x(0)=1, y(0)=0, v_{x}(0)=0, 
v_{y}(0)=1$, I broke a complete orbit into N steps and saw what was 
the resulting error in $y(N)$, which ought to return to zero.  The 
results are shown in the tables below, for various values of N and 
various  values of the source strength g (g=1 gives a circular orbit).

\vskip 0.5 cm
TABLES of computational errors
\vskip 0.5 cm
\begin{tabular}{|l|l|l|l|} \hline
Using $R_{2}$  	& g=0.625 & g=1.0 & g=2.5 \\ \hline
N=100 & 2x$10^{-1}$ & 8x$10^{-3}$ & 2x$10^{-2}$ \\ \hline
N=1,000 & 2x$10^{-3}$ & 8x$10^{-5}$ & 3x$10^{-4}$ \\ \hline
N=10,000 & 2x$10^{-5}$ & 8x$10^{-7}$ & 3x$10^{-6}$ \\ \hline 
\end{tabular}
\vskip 0.5cm
\begin{tabular}{|l|l|l|l|} \hline
Using $R_{4}$  	& g=0.625 & g=1.0 & g=2.5 \\ \hline
N=100 & 3x$10^{-2}$ & 8x$10^{-5}$ & 2x$10^{-3}$ \\ \hline
N=1,000 & 3x$10^{-6}$ & 8x$10^{-9}$ & 2x$10^{-7}$ \\ \hline
N=10,000 & 3x$10^{-10}$ & 8x$10^{-13}$ & 2x$10^{-11}$ \\ \hline 
\end{tabular}
\vskip 1cm

Each increase in the number of 
steps by a factor of 10 improves the accuracy by a factor of $10^{2}$ 
if we use $R_{2}$ and by a factor of $10^{4}$ if we use $R_{4}$. Of 
course, $R _{4}$ requires three times as many operations per step, 
compared to $R_{2}$; but that seems a worthwhile investment since we 
can use many fewer steps for a given overall accuracy.

For comparison, I ran this same calculation using the popular  
 Runge-Kutta method, at second order, and compared the results with those 
shown above for $R_{2}$. Overall, one sees the same rate of improvement in 
accuracy as N is increased; and this is to be expected. For this 
particular problem I found that my method gave somewhat better 
accuracy at each level; but I would not offer that as a general rule 
without much further study; and I encourage others to try both 
methods on their own favorite problems.  I will say, however, that 
the programming for my method was considerably simpler than that for 
the R-K method; and I expect that this aspect of the comparison is 
even more marked as one goes to the fourth order methods.

What about the Richardson technique?  As a general rule, if you calculate something 
with a small parameter $\delta$ and know how it converges to the 
true answer as $\delta \rightarrow 0$, then you can accelerate 
convergence.  For example, if you know
\begin{equation}
A(\delta) = A + \delta^{2}X_{2} + \delta^{4} X_{4} + \ldots,\label{d8}
\end{equation}
then you can do two calculations and combine the results as follows.
\begin{equation}
\frac{4}{3} A(\delta/2) - \frac{1}{3}A(\delta) = A + \delta^{4}Y_{4} 
+ \ldots  .
\end{equation}

I used this method on the Kepler calculation, using $R_{2}$, and 
found results that were slightly better than those obtained from 
using $R_{4}$.  This appears to be a nice alternative  method.

Next I added the Richardson extrapolation to the Runge-Kutta (second 
order) calculation of the same Kepler problem; and found that the 
results were far inferior to those just mentioned.  There was some 
improvement in accuracy but significantly less than expected. The 
reason is that the asymptotic formula (\ref{d8}) is incorrect for the 
Runge-Kutta method: there is a term of order $\delta^{3}$ that 
belongs there.

The main lesson from these experiments appears to be that the 
condition (\ref{b1}) on the lowest order approximation, which we might 
call ``reversibility'', is important.

\section{Velocity-dependent Force}%5

Here we start by considering a simple type of velocity-dependent force:
\begin{equation}
\frac{d^{2}x}{dt^{2}} = f(x,v) = g(x) + v h(x).\label{B1}
\end{equation}
Now, when we construct the approximate propagator $R_{2}(\delta)$, 
the $D_{x}$ 
part will be the same as before but the $D_{v}$ part needs to be changed so as to 
guarantee that $R(-\delta)R(\delta)=1$.

The way we do this is to find the exact solution of the simple 
equation
\begin{equation}
\dot{v} = g + vh,\label{B2}
\end{equation}
where we treat $g=g(x)$ and $h=h(x)$ as constants.  The solution is easy:
\begin{equation}
v(t) = v(0)+ (v(0) + \frac{g}{h})(e^{ht}-1)\label{B3}
\end{equation}
and this shows us how to construct the operator $D_{v}$.

With this special result we can now address the case of a general 
$f(x,v)$. We do this by taking another time-derivative of the 
original equation (\ref{B1}).
\begin{eqnarray}
\dot{x}=v, \;\;\;\;\; \dot{v}=w, \;\;\;\;\; \dot{w} = g(x,v) + w 
h(x,v) \label{B4} \\
g(x,v) = v \frac{\partial f(x,v)}{\partial x}, \;\;\;\;\; 
h(x,v) = \frac{\partial f(x,v)}{\partial v}.\label{B5}
\end{eqnarray}

Now we construct the following.
\begin{eqnarray}
R_{2}(\delta) = 
D_{x}(\delta/2)D_{v}(\delta/2)D_{w}(\delta)D_{v}(\delta/2)D_{x}(\delta/2) \\
D_{x}(\delta) \left(\begin{array}{c} x \\ v \\w \end{array}\right) = 
\left(\begin{array}{c} x+\delta v \\ v \\w \end{array}\right) \\
D_{v}(\delta) \left(\begin{array}{c} x \\ v \\w \end{array}\right) = 
\left(\begin{array}{c} x \\ v +\delta w \\w \end{array}\right) \\
D_{w}(\delta) \left(\begin{array}{c} x \\ v \\w \end{array}\right) = 
\left(\begin{array}{c} x \\ v  \\w +(w+\frac{g}{h})(e^{h\delta}-1)\end{array}\right).
\end{eqnarray}

This technique also shows us how to handle the simplest first order 
equation $\dot{x} = f(x)$ by turning it into a second order equation, 
$\dot{x} = v$ and $\dot{v} = v f'(x)$, and then constructing $R_{2}$ 
as in (\ref{c1}). 
As another option, instead of using
\begin{equation}
D_{v}(\delta): v \rightarrow ve^{\delta f'(x)}
\end{equation}
one could use
\begin{equation}
D_{v}(\delta): v \rightarrow v \frac{1+(\delta/2) 
f'(x)}{1-(\delta/2)f'(x)}.
\end{equation}

\section{Time-dependent Force}%6
Let's return to the original problem  (\ref{a1}) and allow the force to 
be explicitly time dependent.
\begin{equation}
\frac{d}{dt} \left(\begin{array}{c}t\\ x\\v \end{array}\right) = 
\left(\begin{array}{c}1\\ v\\f(t, x) \end{array}\right) = 
M(t) \left(\begin{array}{c}t\\ x\\v \end{array}\right).
\end{equation}

The natural guess is for the second order approximate propagator 
$\tilde{R}_{2}$ to be  built from the original $R_{2}$ as follows.
\begin{equation}
\tilde{R}_{2}(\delta) = D_{t}(\delta/2)R_{2}(\delta)D_{t}(\delta/2), 
\;\;\;\;\; D_{t}(\delta): t \rightarrow t+\delta.
\end{equation}
This means that we proceed as before but evaluate the force $f(t,x)$ 
with the variable $t$ at the midpoint of each time interval. This 
formulation preserves the property (\ref{b1}).

Then we can proceed to construct $\tilde{R}_{4}$ just as before, 
using three of these operators  $\tilde{R}_{2}$ with the same weight 
factors $\beta, \gamma$ as in 
(\ref{b4}). 

\section{Discussion}%7

The method described here appears to be a powerful, simple and versatile tool. 

It should be apparent how one can continue to improve the method, 
going from $R_{4}$ to $R_{6}$ , etc. While it is not easy to guess in 
advance what level of accuracy will be most efficient in any given 
problem, the programming procedures outlined above make it relatively 
easy to experiment and find the best approach.

It is interesting that the equations (\ref{b6}) have another solution, one 
that goes into the complex plane, as follows.
\begin{equation}
\beta \approx 0.324 \pm 0.135 i, \;\;\;\;\; \gamma \approx 0.352 \mp 0.270 i.
\end{equation}
Which is the best to use?  My guess is that for conservative systems, 
the real solutions are best but for dissipative systems this complex 
scheme may be better. This whole area needs further study and 
experimentation.

\vskip 0.5cm
\setcounter{equation}{0}
\def\theequation{A.\arabic{equation}}
\boldmath
\noindent{\bf Appendix: Many dimensions and many particles}%A
\unboldmath
\vskip 0.5cm

If we have just one particle moving in several dimensions, all we need to do is replace the single quantities $x,v, f(x)$ in the previous equations by vector quantities $\textbf{x}, \textbf{v}, \textbf{f}$.

If we have N particles, labeled by $ i = 1, ..., N$, then something more needs to be done.  Let's write $R_2(i, \delta)$ as the process, like that shown in Section 3, for advancing coordinates $\textbf{x}_i, \textbf{v}_i$ by the time interval $\delta$, while keeping all other particle coordinates fixed. That means we need a separate subroutine $\textbf{f}(i)$ that calculates the force acting on the ith particle given the positions (and velocities) of all the other particles. Now we construct a generalized "sandwich of operations", which is easiest written as two lines of computer code (in the C language).
\begin{eqnarray}
for(i=1;i<=N; i++) R_2(i,\delta/2); \\
for(i=N; i>=1; i--) R_2(i, \delta/2);
\end{eqnarray}
This is the full operation of $R_2 (\delta)$ for the whole system; and it preserves the reversible property $R(-\delta)R(\delta) = 1$. We can then go on to $R_4(\delta) = R_2 (\beta \delta)R_2 (\gamma \delta)R_2(\beta\delta)$, as before

If we have velocity-dependent forces in several dimensions, the method shown in Section 5 needs modification. Again, the neat answer is a sandwich process that advances one component of the velocity at each step, with all the other components kept fixed.

\end{document}